\documentclass[twocolumn,twoside,slac_two]{revtex4}
\usepackage{graphicx}
\usepackage{fancyhdr}
\newcommand{\ber}{\mbox{$^{7}$Be} }

\newcommand{\cele}{\mbox{$^{11}$C} }

\newcommand{\ctwo}{\mbox{$^{12}$C} }

\newcommand{\kr}{\mbox{$^{85}$Kr} }

\newcommand{\rbm}{\mbox{$^{\rm 85m}$Rb} }
\newcommand{\bite}{\mbox{$^{210}$Bi} }
\newcommand{\potwoten}{\mbox{$^{210}$Po} }
\newcommand{\ledtwoten}{\mbox{$^{210}$Pb} }

\newcommand{\Bipo}{\mbox{$^{214}$Bi-$^{214}$Po} }

\newcommand{\tho}{\mbox{$^{232}$Th} }

\newcommand{\ura}{\mbox{$^{238}$U} }
\newcommand{\che}{\v{C}erenkov }

\newcommand{\nue}{\mbox{$\nu_e$} }
\newcommand{\num}{\mbox{$\nu_\mu$} }

\begin{document}

\title{The first year of Borexino}

\newcommand{\APC}{Laboratoire AstroParticule et Cosmologie, 75231 Paris cedex 13, France}
\newcommand{\Budapest}{KFKI-RMKI, 1121 Budapest, Hungary}
\newcommand{\Dubna}{Joint Institute for Nuclear Research, 141980 Dubna, Russia}
\newcommand{\Genova}{Dipartimento di Fisica, Universit\`a e INFN, Genova 16146, Italy}
\newcommand{\Heidelberg}{Max-Planck-Institut f\"ur Kernphysik, 69029 Heidelberg, Germany}
\newcommand{\Kiev}{Kiev Institute for Nuclear Research, 06380 Kiev, Ukraine}
\newcommand{\Krakow}{M.~Smoluchowski Institute of Physics, Jagiellonian University, 30059 Krakow, Poland}
\newcommand{\Kurchatov}{RRC Kurchatov Institute, 123182 Moscow, Russia}
\newcommand{\LNGS}{INFN Laboratori Nazionali del Gran Sasso, SS 17 bis Km 18+910, 67010 Assergi (AQ), Italy}
\newcommand{\Milano}{Dipartimento di Fisica, Universit\`a degli Studi e INFN, 20133 Milano, Italy}
\newcommand{\Munich}{Physik Department, Technische Universit\"at Muenchen, 85747 Garching, Germany}
\newcommand{\Pavia}{INFN, Pavia 27100, Italy}
\newcommand{\Perugia}{Dipartimento di Chimica, Universit\`a e INFN, 06123 Perugia, Italy}
\newcommand{\Peters}{St. Petersburg Nuclear Physics Institute, 188350 Gatchina, Russia}
\newcommand{\Princeton}{Physics Department, Princeton University, Princeton, NJ 08544, USA}
\newcommand{\PrincetonChemEng}{Chemical Engineering Department, Princeton University, Princeton, NJ 08544, USA}
\newcommand{\Queens}{Physics Department, Queen's University, Kingston ON K7L 3N6, Canada}
\newcommand{\UMass}{Physics Department, University of Massachusetts, Amherst, MA 01003, USA}
\newcommand{\Virginia}{Physics Department, Virginia Polytechnic Institute and State University, Blacksburg, VA 24061, USA}
\newcommand{\df}{\thanks{\mbox{Contribution Presented by D. Franco.} \mbox{\emph{Email: } Davide.Franco@mi.infn.it}}}

\author{D.~Franco}\df\affiliation{\Milano}
\author{G.~Bellini}\affiliation{\Milano}
\author{J.~Benziger}\affiliation{\PrincetonChemEng}
\author{S.~Bonetti}\affiliation{\Milano}
\author{M.~Buizza Avanzini}\affiliation{\Milano}
\author{B.~Caccianiga}\affiliation{\Milano}
\author{L.~Cadonati}\affiliation{\UMass}
\author{F.~Calaprice}\affiliation{\Princeton}
\author{C.~Carraro}\affiliation{\Genova}
\author{A.~Chavarria}\affiliation{\Princeton}
\author{F.~Dalnoki-Veress}\affiliation{\Princeton}
\author{D.~D'Angelo}\affiliation{\Milano}
\author{H.~de~Kerret}\affiliation{\APC}
\author{A.~Derbin}\affiliation{\Peters}
\author{A.~Etenko}\affiliation{\Kurchatov}
\author{K.~Fomenko}\affiliation{\Dubna}
\author{C.~Galbiati}\affiliation{\Princeton}
\author{S.~Gazzana}\affiliation{\LNGS}
\author{M.~Giammarchi}\affiliation{\Milano}
\author{M.~Goeger-Neff}\affiliation{\Munich}
\author{A.~Goretti}\affiliation{\Princeton}
\author{C.~Grieb}\altaffiliation{Present address: European Patent Office, M\"unich, Germany.}\affiliation{\Virginia}
\author{S.~Hardy}\affiliation{\Virginia}
\author{Aldo~Ianni}\affiliation{\LNGS}
\author{Andrea~Ianni}\affiliation{\Princeton}
\author{M.~Joyce}\affiliation{\Virginia}
\author{V.~Kobychev}\affiliation{\Kiev}
\author{G.~Korga}\affiliation{\LNGS}
\author{D.~Kryn}\affiliation{\APC}
\author{M.~Laubenstein}\affiliation{\LNGS}
\author{M.~Leung}\affiliation{\Princeton}
\author{T.~Lewke}\affiliation{\Munich}
\author{E.~Litvinovich}\affiliation{\Kurchatov}
\author{B.~Loer}\affiliation{\Princeton}
\author{P.~Lombardi}\affiliation{\Milano}
\author{L.~Ludhova}\affiliation{\Milano}
\author{I.~Machulin}\affiliation{\Kurchatov}
\author{S.~Manecki}\affiliation{\Virginia}
\author{W.~Maneschg}\affiliation{\Heidelberg}
\author{G.~Manuzio}\affiliation{\Genova}
\author{F.~Masetti}\affiliation{\Perugia}
\author{K.~McCarty}\affiliation{\Princeton}
\author{Q.~Meindl}\affiliation{\Munich}
\author{E.~Meroni}\affiliation{\Milano}
\author{L.~Miramonti}\affiliation{\Milano}
\author{M.~Misiaszek}\affiliation{\Krakow}\affiliation{\LNGS}
\author{D.~Montanari}\affiliation{\LNGS}\affiliation{\Princeton}
\author{V.~Muratova}\affiliation{\Peters}
\author{L.~Oberauer}\affiliation{\Munich}
\author{M.~Obolensky}\affiliation{\APC}
\author{F.~Ortica}\affiliation{\Perugia}
\author{M.~Pallavicini}\affiliation{\Genova}
\author{L.~Papp}\affiliation{\LNGS}
\author{L.~Perasso}\affiliation{\Milano}
\author{S.~Perasso}\affiliation{\Genova}
\author{A.~Pocar}\altaffiliation{Now at Stanford University, CA, USA.}\affiliation{\Princeton}
\author{R.S.~Raghavan}\affiliation{\Virginia}
\author{G.~Ranucci}\affiliation{\Milano}
\author{A.~Razeto}\affiliation{\LNGS}
\author{P.~Risso}\affiliation{\Genova}
\author{A.~Romani}\affiliation{\Perugia}
\author{D.~Rountree}\affiliation{\Virginia}
\author{A.~Sabelnikov}\affiliation{\Kurchatov}
\author{R.~Saldanha}\affiliation{\Princeton}
\author{C.~Salvo}\affiliation{\Genova}
\author{S.~Sch\"onert}\affiliation{\Heidelberg}
\author{H.~Simgen}\affiliation{\Heidelberg}
\author{M.~Skorokhvatov}\affiliation{\Kurchatov}
\author{O.~Smirnov}\affiliation{\Dubna}
\author{A.~Sotnikov}\affiliation{\Dubna}
\author{S.~Sukhotin}\affiliation{\Kurchatov}
\author{Y.~Suvorov}\affiliation{\Milano}\affiliation{\Kurchatov}
\author{R.~Tartaglia}\affiliation{\LNGS}
\author{G.~Testera}\affiliation{\Genova}
\author{D.~Vignaud}\affiliation{\APC}
\author{R.B.~Vogelaar}\affiliation{\Virginia}
\author{F.~von~Feilitzsch}\affiliation{\Munich}
\author{M.~Wojcik}\affiliation{\Krakow}
\author{M.~Wurm}\affiliation{\Munich}
\author{O.~Zaimidoroga}\affiliation{\Dubna}
\author{S.~Zavatarelli}\affiliation{\Genova}
\author{G.~Zuzel}\affiliation{\Heidelberg}

\begin{abstract}

Borexino is an experiment designed to detect in real-time low energy solar neutrinos. 
It is installed at the Gran Sasso Underground Laboratory and has started taking data
in May 2007.  We report the direct measurement of the 
\ber\ solar neutrino signal rate after 1 year of data taking. 
Implications and perspectives are discussed.

\end{abstract}

\maketitle

\thispagestyle{fancy}

\section{Introduction}

The Sun is an intense  source of electron neutrinos, produced
in the nuclear reactions of the proton-proton chain and of the CNO cycle. 
Solar neutrinos provide a unique probe for studying both the nuclear fusion reactions that power the
Sun and the fundamental properties of neutrinos. 

In particular, neutrino oscillations, described in the  Large Mixing Angle (LMA) Mykheyev-Smirnov-Wolfenstein
(MSW)~\cite{MSW,Abe08} theory,   offer a solution to the solar neutrino problem, the long standing
discrepancy between the observation of the solar neutrino flux in the pioneer radiochemical and
water Cherenkov experiments~\cite{SNP} and the prediction of the Standard Solar Model~\cite{BS07}.

A central feature of the LMA-MSW solution is the prediction that neutrino oscillations are dominated by
vacuum oscillations at low energies ($<$2~MeV) and by resonant matter-enhanced oscillations, taking place
in the Sun's core, at higher energies ($>$5~MeV).  

Borexino is the first experiment to report a real-time observation of low energy solar neutrinos in the
vacuum oscillation regime  by the direct measurement of the low energy (0.862 MeV) \ber\ solar neutrino
interaction rate.  We report here the results \cite{BX07,BX08} from an analysis of 192 live days of Borexino detector
live-time in the period from May 16, 2007 to April 12, 2008, totaling a 41.3 ton yr fiducial exposure to
solar neutrinos.

\begin{figure} 
\centering
\includegraphics[width=80mm]{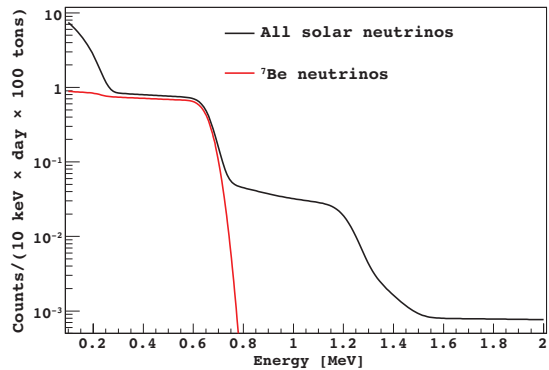}
\caption{Neutrino spectra expected in Borexino (accounting
for the detector's energy resolution). The solid black line
represents the neutrino signal rate in Borexino according to
the most recent predictions of the Standard Solar Model \cite{BS07}
including neutrino oscillations with the LMA-MSW parameters \cite{Abe08}.
The solid red line illustrates the contribution due to
\ber\ neutrinos. pp neutrinos contribute to the spectrum below
0.3 MeV and the edge at 1.2 MeV is due to p-e-p neutrinos.} 
\label{fig:nu}
\end{figure}

Solar neutrinos are detected in Borexino through their elastic scattering on electrons in the scintillator. Electron neutrinos (\nue)
interact through charged and neutral currents and in the energy range of interest have a cross section 5 times larger than \num and
$\nu_\tau$, which interact only via the neutral current. The electrons scattered by neutrinos are detected by means of the scintillation light
retaining the information on the energy, while information on the direction of the scattered electrons is lost. The basic signature for
the mono-energetic 0.862 MeV \ber neutrinos is the Compton-like edge of the recoil electrons at 665 keV, as shown in Figure
\ref{fig:nu}.

A strong effort has been devoted to the containment and comprehension of the background,
since electron-like events induced by solar neutrino interactions can not be distinguished, on
an event-by-event basis, from electrons or photons due to radioactive decays. 
The main challenge is then to reduce the background in the active mass in order to reach a signal-to-noise
ratio equal to 1. The detector has been built following a self
shielding design by increasing the
radiopurity requirements while moving closer to the active mass. In order to reach a signal-to-noise
ratio of 1 an intrinsic radiopurity of  4 10$^{-4}$$\mu$Bq/kg is required both for
\ura\ and \tho\ contaminations. The Borexino purification strategy relies on
filtration at the level of 0.05 $\mu$m, multi-stage distillation and high purity nitrogen
sparging.

\section{The Detector}

\begin{figure} 
\centering
\includegraphics[width=80mm]{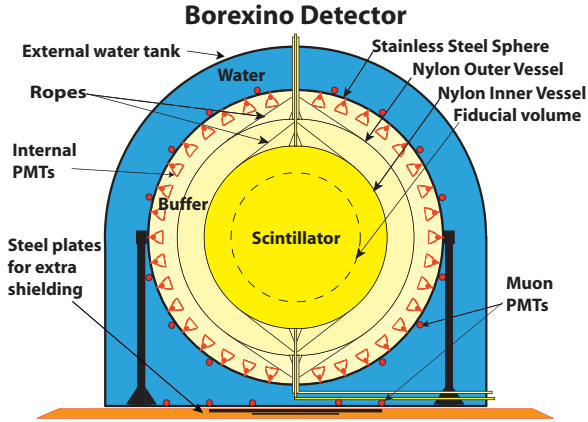}
\caption{Schematic drawing of the Borexino detector.} \label{fig:scheme}
\end{figure}

The Borexino detector is located at the Gran Sasso National Laboratories (LNGS) in central Italy, at a depth of 3800~m.w.e..
Neutrinos are detected via elastic scattering off electrons in liquid scintillator. The sketch of the detector is shown in Figure
\ref{fig:scheme}. The active target consists of 278~tons of pseudocumene (PC, 1,2,4 trimethylbenzene), doped with 1.5~g/liter of PPO
(2,5-diphenyloxazole, a fluorescent dye). The scintillator is contained in a thin (125~$\mu$m) nylon vessel and is shielded by two
concentric PC buffers (323 and 567~tons) doped with 5.0~g/l of a scintillation light quencher (dimethylphthalate). The two PC buffers
are separated by a second thin nylon membrane to prevent diffusion of radon towards the scintillator. The scintillator and buffers
are contained in a Stainless Steel Sphere (SSS) with a diameter of 13.7~m. The SSS is enclosed in a 18.0-m diameter, 16.9-m high
domed Water Tank (WT), containing 2100~tons of ultra-pure water as an additional shield. The scintillation light is detected via 2212
8" photomultiplier tubes (PMTs) uniformly distributed on the inner surface of the SSS. Additional 208 8" PMTs instrument the WT and
detect the Cherenkov light radiated by muons in the water shield, serving as a muon veto. A detailed description of the detector can
be found in~\cite{BXD08}.

An {\it event} in Borexino is recorded when at least 25 PMT pulses occur within
a time window of 99 ns (the corresponding energy threshold is about 40 keV). 
When a trigger occurs, a 16 $\mu$s gate is opened and time and charge of each PMT
pulse is collected. The offline software identifies the shape and the length of each scintillation
pulse and reconstructs the position of the energy deposit in the scintillator by means of a time of
flight technique. Pulse shape analysis is performed to identify various classes of events, among which
electronic noise, pile up events, muons, $\alpha$ and $\beta$ particles.

\section{Event Selection and Spectral Fits}

The analyzed energy range is 250--800 keV. The event selection relies on the following cut criteria:

\vskip0.1in

\begin{figure} 
\centering
\includegraphics[angle=90,width=80mm]{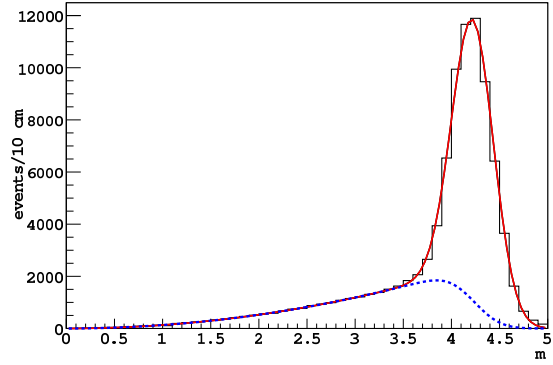}
\caption{Fit of the radial distribution of the events in the $^{7}$Be energy region. The fit function has two components: bulk events (blue) and events from the
nylon vessel and from the buffer (red). The fiducial volume is defined with a radial cut at 3 m. } \label{fig:radial}
\end{figure}

\begin{enumerate}
\item The event must have a unique reconstructed cluster of PMT hits, in order to reject pile-up events and fast coincident events in the same
acquisition window.  The efficiency of this cut is nearly~100\% because the very low triggering rate results in a negligible pile-up.
\item Events with \che\ light in the water tank detector are identified as cosmic muons and rejected.
\item All the detector is vetoed for 2~ms  after each  muon crossing the scintillator. In this way, muon-induced neutrons (mean capture time 
$\sim$ 250 $\mu$s) and spurious events like after-pulses are rejected. The measured muon rate in Borexino (muons that cross the scintillator and buffer volume) is 0.055$\pm$0.002~s$^{-1}$.  
The dead time introduced by this cut is negligible.
\item Decays due to radon daughters occurring before the \Bipo\ delayed coincidences are vetoed. The fraction surviving the veto is accounted for in the analysis.
\item The events must be reconstructed within a spherical fiducial volume corresponding nominally to 100~ton in order to reject external~$\gamma$~background
(Figure \ref{fig:radial}).  
Another volumetric cut ($z$$<$1.8~m) was applied in order to remove a small background from~$^{222}$Rn~daughters in the north pole of the inner vessel, 
resulting in a nominal fiducial mass of 87.9 t. 
\end{enumerate}

\begin{figure} 
\centering
\includegraphics[width=80mm]{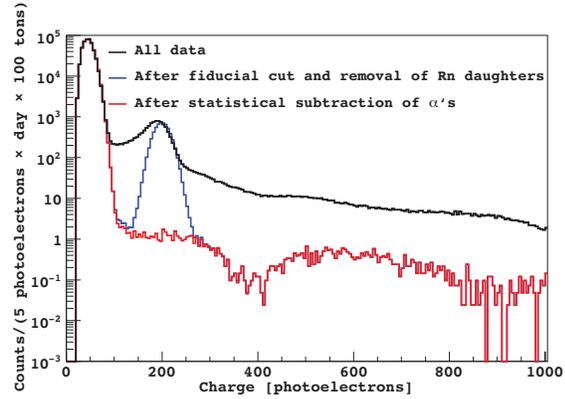}
\caption{The raw photoelectron charge spectrum after the
cuts 1-3 (black), after the fiducial cut 5 (blue), and after
the statistical subtraction of the $\alpha$-emitting contaminants
(red). All curves scaled to the exposure of 100 day ton. Cuts are
described in the text.} \label{fig:cuts}
\end{figure}

\begin{figure} 
\centering
\includegraphics[width=80mm]{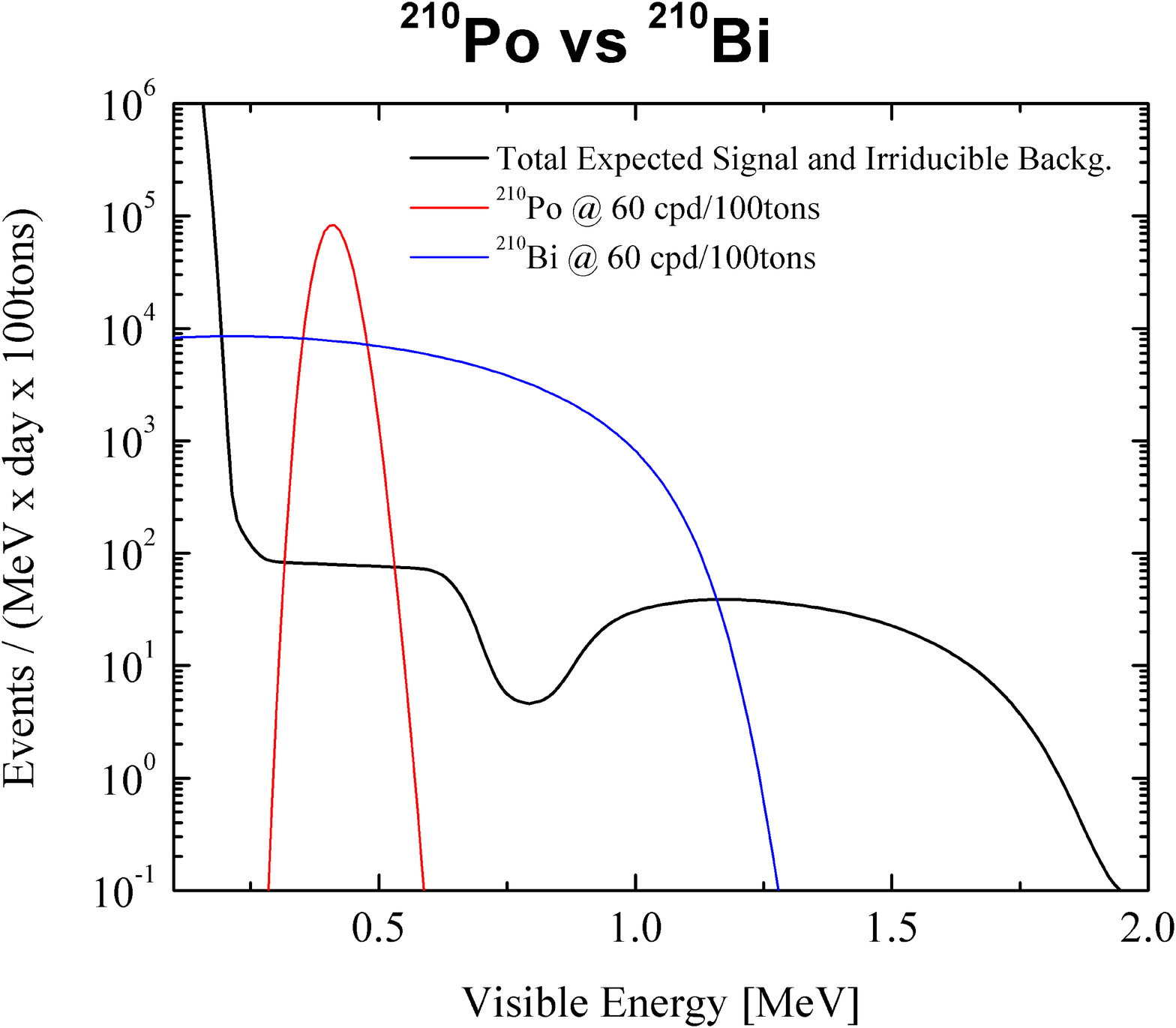}
\caption{Expected \bite contribution in case of \ledtwoten contamination.} \label{fig:bi210}
\end{figure}

\begin{figure} 
\centering
\includegraphics[width=80mm]{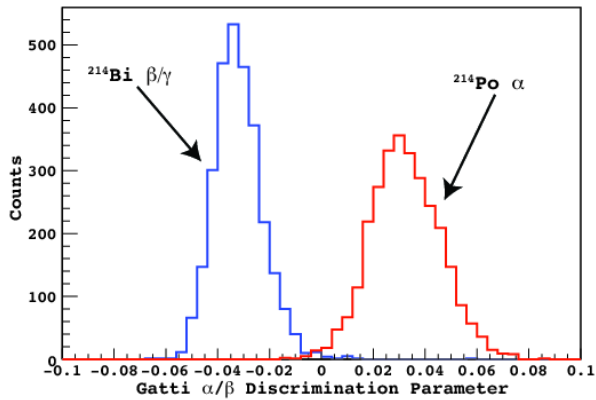}
\caption{$\alpha$/$\beta$ discrimination of $^{214}$Bi-$^{214}$Po events with the Gatti filter.} \label{fig:ab}
\end{figure}

In Figure \ref{fig:cuts} the measured spectrum in 192 days is shown before and after the cuts. It can be noticed that an important  peak is present in the
data after the fiducial volume cut. This peak is due to a \potwoten contamination still present in the liquid scintillator after purification and filling. 
\potwoten is produced  in decay chain segment starting from $^{210}$Pb. \ledtwoten decays to \bite which has an end-point energy at about 1 MeV. The measured activity of \potwoten 
is clearly not in
equilibrium with a \bite source, as shown in Figure \ref{fig:bi210}. Moreover, the measured decay trend has a mean life compatible with $^{210}$Po.  
Even if \ledtwoten contamination is high with respect to the expected signal, it can be efficiently identified in the offline analysis with the pulse shape
discrimination, shown in Figure \ref{fig:ab}.  A filter based on the time response of the PC-based scintillator \cite{BX082}, slower for $\alpha$ particles 
than for $\beta$'s, discriminates the $^{210}$Po events. 
The red curve in Figure \ref{fig:cuts} shows the effect of the statistical subtraction of
the $\alpha$-emitting contaminants, by use of the pulse shape discrimination.

\begin{figure} 
\centering
\includegraphics[width=80mm]{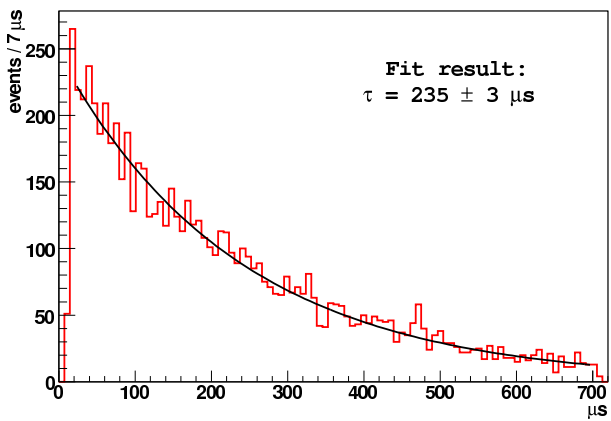}
\caption{Time difference between the first and second event
of the $^{214}$Bi-$^{214}$Po coincidence sample. The resulting lifetime from the fit (235$\pm$3 $\mu$s) is in agreement with the $^{214}$Po mean life (237 $\mu$s)} \label{fig:bipo}
\end{figure}

The study of fast coincidence decays of $^{214}$Bi-$^{214}$Po (see Figure \ref{fig:bipo})
 and $^{212}$Bi-$^{212}$Po, yields, under
the assumption of secular equilibrium, to the estimation of  \ura contamination equal to (1.6$\pm$0.1)10$^{-17}$ g/g and of \tho equal to
(6.8$\pm$1.5)10$^{-18}$ g/g. The \kr content in the scintillator
was probed by looking the delayed coincidence in the secondary branch of \kr decay throughout the metastable level \rbm (BR =0.43\%).
Our best estimate for the activity of \kr is 29$\pm$14 counts/(day$\cdot$100 ton).

From the spectrum in Figure \ref{fig:cuts}  the expected Compton-like edge due to \ber solar neutrinos is well visible. 
Moreover, it can be seen that at high energy the spectrum is
dominated by a cosmogenic component well known, the \cele. \cele is produced underground by muons interacting with \ctwo 
in the liquid scintillator. This background depends on the depth of the underground laboratory \cite{Gal05,BX06}. 

\begin{figure} 
\centering
\includegraphics[width=80mm]{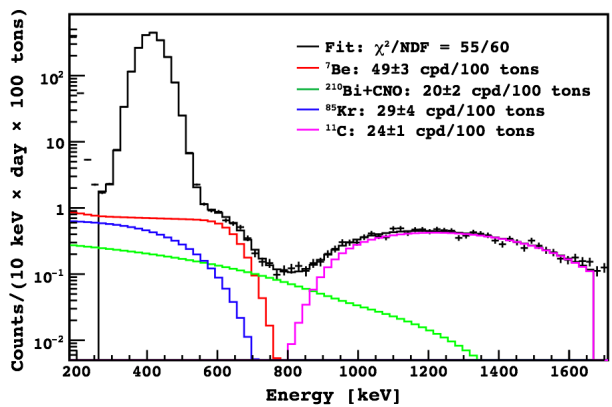}
\caption{Spectral fit in the energy region 260-1670 keV before the $\alpha$ subtraction.} \label{fig:fit1}
\end{figure}

\begin{figure} 
\centering
\includegraphics[width=80mm]{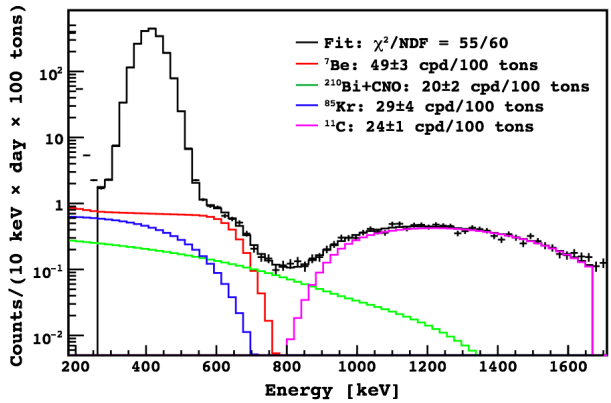}
\caption{Spectral fit in the energy region 160-2000 keV.} \label{fig:fit2}
\end{figure}

The spectra within the fiducial volume was studied with and without the $\alpha$ statistical subtraction. Results are shown in
Figures \ref{fig:fit1} and \ref{fig:fit2}. In the spectral fits, the contribution of CNO neutrinos is combined with
that of \bite which is not known. The two spectra are degenerate in the \ber
region. The \ber, the \kr, the \cele as well as the light yield are free parameters
of the fit. A light yield of about 500 p.e./MeV is found for $\beta$'s, and the energy
resolution scales approximately as 5\%/$\sqrt{E[MeV]}$

Systematic uncertainties come mainly from the total scintillator mass
(0.2\%), the fiducial mass definition (6\%) and the detector response function (6\%).

Taking into account systematic errors, our best value for the interaction rate of the 0.862 MeV \ber solar neutrinos is 49$\pm$3 (stat) $\pm$ 4 (syst)
counts/(day100 ton). 
The corresponding flux solar neutrino flux $\Phi$(\ber)=(5.08$\pm$0.25)x10$^9$ cm$^{-2}$ s$^{-1}$ is evaluated assuming the 
oscillation
parameters, sin$^2$2$\theta_{12}$ =0.87 and  $\Delta$m$_{12}^2$ =7.6x10$^{-5}$  eV$^2$, from  \cite{Abe08}, in good agreement with expectations.

\section{Results and Perspectives}

The expected neutrino interaction rate in case of no oscillations is 74$\pm$4 counts/(day100 ton), 
The Borexino measurement of the $^7$Be neutrino rate confirms the oscillation hypothesis at 4 $\sigma$, in agreement with the MSW-LAM prediction.

Under the assumption of the SSM constraint the solar neutrino survival probability is measured to be Pee = 0.56 $\pm$ 0.10.

The Borexino measured rate can be combined with the other solar neutrino measurements
to constrain the  flux normalization constants of the other solar neutrino fluxes. This leads to  the best determination
of the pp solar neutrinos flux, obtained with the assumption of the luminosity constraint:  $f_{pp}=1.005^{+0.008}_{-0.020}$
where f$_{pp}$ is the ratio between the measured and predicted pp neutrino fluxes. 
With the same technique, Borexino obtained the best limit on the CNO flux: $f_{CNO}<6.27$ (90\% C.L.). 

The low energy solar neutrino spectrum is sensitive to the possible presence of a non-null magnetic moment. We exploited this feature 
to determine the best upper limit to the neutrino magnetic moment  (5.4x10$^{-11}$  $\mu$B, 90\% C.L.) \cite{BX08}.

The first Borexino results have shown for the first time the feasibility to measure
solar neutrinos in the sub-MeV range in real-time. Moreover, the high level
of radiopurity achieved allows to investigate other solar neutrino sources. In particular,  
CNO and p-e-p  neutrino  detections depend on the possibility to tag and reject event by event 
cosmogenic \cele background. This goal is at present under investigation, since \cele can be identified by means of the
 three-fold coincidence with the parent muon and  the following 
neutron emission and capture on hydrogen. 

Thanks to the excellent scintillator radio-purity, Borexino has also the opportunity to measure
the $^8$B neutrino spectrum  with the lowest energy threshold so far. A dedicated analysis with the energy threshold at 2.8 MeV is in progress.


\begin{thebibliography}{9}   
\bibitem{MSW} S.P.~Mikheev and A.Yu.~Smirnov, Sov. J. Nucl. Phys. {\bf 42}, 913 (1985);
L.~Wolfenstein, Phys. Rev. D {\bf 17}, 2369 (1978);
P.C.~de~Holanda and A.Yu.~Smirnov, JCAP 0302, 001 (2003).
\bibitem{Abe08} S. Abe et al., (KamLAND Collaboration), arXiv:0801.4589v2, submitted to Phys. Rev. Lett. (2008). 
\bibitem{SNP} B.T.~Cleveland et al., Ap. J. {\bf 496}, 505 (1998);
K.~Lande and P.~Wildenhain, Nucl. Phys. B (Proc. Suppl.) {\bf 118}, 49 (2003);
R.~Davis, Nobel Prize Lecture (2002).
W.~Hampel et al. (GALLEX Collaboration), Phys. Lett. B {\bf 447}, 127 (1999);
J.N.~Abdurashitov et al. (SAGE collaboration), Phys. Rev. Lett. {\bf 83}, 4686 (1999);
M.~Altmann et al. (GNO Collaboration), Phys. Lett. B {\bf 616}, 174 (2005);
K.S.~Hirata et al. (KamiokaNDE Collaboration), Phys. Rev. Lett. 63, 16 (1989).
\bibitem{BS07} J.N.~Bahcall, A.M.~Serenelli, and S.~Basu, Astrophys. J. Suppl. {\bf 165}, 400 (2006);
C.~Pe\~na-Garay, talk at the conference Neutrino Telescopes 2007", March 6-9, 2007, Venice,
{\tt http://neutrino.pd.infn.it/conference2007/}; A.~Serenelli, private communication.
\bibitem{BX07} C.~Arpesella et al. (Borexino Collaboration), Phys. Lett. B {\bf 658}, 101 (2007).
\bibitem{BX08} C.~Arpesella et al. (Borexino Collaboration), accepted for publication on Phys. Rev. Lett., arXiv:0805.3843 (2008).

\bibitem{BXD08} G.~Alimonti et al. (Borexino Collaboration), submitted to Nucl. Instr. Meth. A, arXiv:0806.2400 (2008).
\bibitem{BX082} H.O. Back et al.(Borexino Collaboration),  Nucl. Instrum. Meth. A {\bf 584} 98-113 (2008); 
\bibitem{Gal05} C.~Galbiati et al., Phys.Rev.C {\bf 71} 055805 (2005).
\bibitem{BX06} H.~Back et al. (Borexino Collaboration), Phys. Rev. C {\bf 74}, 045805 (2006).



\end{thebibliography}
\end{document}